\documentclass[sigconf]{acmart}
\setcopyright{none}
\renewcommand\footnotetextcopyrightpermission[1]{}
\begin{document}

\title{Toward System-of-Systems Integration for Composable Cloud-HPC-Edge AI Platforms}
\author{Sumit Rakesh }
\email{Sumit.Rakesh@Ltu.se }
\affiliation{%
  \institution{Lule\aa~tekniska universitet}
  \city{Lule\aa}
  \country{Sweden}
}


\author{Rajkumar Saini}
\email{Rajkumar.Saini@Ltu.se }
\affiliation{%
  \institution{Lule\aa~tekniska universitet}
  \city{Lule\aa}
  \country{Sweden}
}

\renewcommand{\shortauthors}{Rakesh et al.}

\begin{abstract}
Modern AI platforms increasingly combine infrastructure stacks and operating models designed around different assumptions, including cloud-style service platforms, HPC workload-management systems, cloud-native orchestration, data and artifact systems, managed connectivity, observability, and edge or cyber-physical environments. Existing work demonstrates effective bridges between selected stacks, but a general way to reason about composition across independently controlled systems remains underdeveloped. We argue that such platforms can be usefully viewed as systems of systems (SoS) when independently useful systems retain their own control, management, lifecycles, policies, and failure semantics while contributing to a higher-level AI platform capability. We frame composable integration as an approach to cross-system coordination based on interfaces, contracts, mappings, references, policy context, and operational evidence, while preserving native control planes and avoiding dependence on a single topology or orchestration stack. The resulting direction is converged in use and federated in control. The paper presents a peer constituent-system view, a boundary test for distinguishing constituent systems from components, local dependencies, and independently useful systems outside the current SoS boundary, a local, shared, and scoped responsibility model, seven integration surfaces, and a representative cross-system workflow. It concludes with evidence classes and research questions for evaluating interoperability, governance, observability, fault containment, evolution, and reuse.
\end{abstract}



\maketitle
\section{Introduction}

AI infrastructure increasingly spans operating domains designed around different assumptions. A contemporary AI platform may combine large-scale training and fine-tuning, interactive development, online and batch inference, governed data and artifact management, accelerator-aware execution, cross-system observability, and interaction with sensors, cameras, robots, immersive clients, or industrial gateways. Cloud-style service platforms emphasize accessible services and resource abstraction \cite{1.mell2011nist}. HPC systems emphasize managed execution, accelerator locality, high-performance communication, and accounting \cite{2.schedmdSlurmOverview}. Cloud-native orchestration emphasizes declarative control, automation, and repeatable deployment \cite{3.cncfCloudNativeDefinition,4.kubernetesComponents}. Edge and cyber-physical systems emphasize locality, latency, device ownership, intermittent connectivity, and safety or privacy constraints \cite{5.etsiMec003,19.belcastro2026navigating,20.rosendo2022distributed}. The integration challenge emerges when these systems must participate in shared AI workflows while retaining their distinct control and operating models. 

Whether a Cloud-HPC-Edge AI platform should be treated as a system of systems (SoS) depends on the degree of independence among its participating systems. The SoS framing becomes appropriate when independently useful systems retain their own control, management, lifecycle, and failure semantics while contributing to a higher-level capability that no single system can provide alone. Although a multi-site federation makes the SoS character readily apparent, a single-site deployment may exhibit the same properties when independently managed systems retain distinct operators, lifecycles, policies, and control planes. Maier identifies operational and managerial independence as defining SoS characteristics \cite{6.maier1998architecting}, while ISO/IEC/IEEE guidance treats constituent systems as systems with their own lifecycle concerns that also participate in a SoS \cite{7.iso21839,8.iso21840}. These properties provide a stronger basis for defining constituent-system boundaries than grouping technologies within the same deployment. 

We present composable integration as the corresponding approach to SoS integration. It is distinct from composable infrastructure in the hardware-disaggregation sense and from application-service composition through a service mesh. Prior composable-system work focuses on assembling compute, memory, storage, or accelerator resources under an infrastructure control plane \cite{9.elmaghraoui2021composable}, while service-mesh guidance addresses communication and policy among application services \cite{10.chandramouli2020serviceMesh}. Our concern is platform-level composition across independently controlled systems. Such a composition relies on explicit interfaces, workflow contracts, identity and policy mappings, dataset and artifact references, deployment specifications, telemetry, and recovery expectations rather than on replacing native control planes or imposing a single orchestration stack. This paper makes three conceptual contributions. 

\indent 1. First, it introduces a peer constituent-system integration model and a boundary test that distinguishes constituent systems from components, local dependencies, and independently useful systems outside the current SoS boundary. 

\indent 2. Second, it defines local, shared, and scoped responsibilities together with seven integration surfaces that preserve native authority while making cross-system obligations explicit. 

\indent 3. Third, it maps a representative remotely initiated edge-to-training-to-serving workflow across these views and derives testable evidence and research questions concerning interoperability, governance, fault containment, evolution, and reuse. 

\section{Cloud-HPC-Edge Convergence and the Architectural Gap}

Three trends make the problem immediate. First, AI workload and accelerator diversity are widening. Training and inference suites cover workloads with different latency, data, memory, accuracy, and throughput objectives \cite{11.mlcommonsTraining,12.reddi2020mlperf}. Hardware surveys likewise span GPUs, TPUs, FPGAs, ASICs, NPUs, RISC-V accelerators, near-memory designs, and other co-processors \cite{13.silvano2025survey}. Placement, therefore, depends on workload stage, data locality, latency, accelerator capability, energy envelope, trust boundary, and failure risk---not only available compute.

Second, the Cloud-HPC divide is becoming operationally visible. AI-factory work argues for a dual-stack direction that combines HPC performance with cloud-native usability and service-facing interfaces \cite{14.garcialopez2025aifactories}. Current AI-factory reference architectures likewise separate compute, storage, external access, support, and out-of-band management network roles, illustrating the distinct operational paths present even within tightly integrated deployments \cite{34.nvidia2026networking}. Systems research has explored container orchestration on HPC \cite{15.zhou2020container}, cloud-native workloads on HPC resources while preserving HPC accounting \cite{16.chazapis2024cloudnative}, multi-tenant RDMA for Kubernetes on HPC fabrics \cite{17.friese2025closing}, and Kubernetes-Slurm integration for accelerator-backed LLM serving \cite{18.trappen2025automated}. These results show that selected bridges are feasible; they do not, by themselves, define ownership, policy, storage exposure, telemetry, or recovery across the complete platform.

Third, AI increasingly spans the edge-to-cloud continuum. Surveys emphasize decentralized resources, low-latency processing, IoT-generated data, heterogeneous deployment models, and reproducibility challenges \cite{19.belcastro2026navigating,20.rosendo2022distributed}. DECICE adds scheduling, monitoring, and digital-twin coordination across cloud, HPC, and edge environments \cite{21.sharma2026decice}. Research infrastructures such as the National Research Platform and CHI@Edge demonstrate multi-site Kubernetes, heterogeneous GPU/CPU resources, portals, edge experimentation, and operational monitoring \cite{22.weitzel2025national,23.keahey2025chiedge}. EuroHPC coordination of AI Factories and Antennas similarly targets technical and procedural interoperability and the sharing of data, applications, services, and knowledge across a distributed AI ecosystem \cite{35.eurohpc2026aifactories}. Scientific cyberinfrastructure work documents convergence between AI and HPC, similarly at scale \cite{24.huerta2020convergence}.

Existing references clarify important slices of the problem. HPC security guidance defines specialized threat and architecture concerns \cite{25.guo2024hpcsecurity}, while Zero Trust makes access decisions resource-centric rather than location-centric \cite{26.rose2020zerotrust}. Kubernetes multi-tenancy guidance distinguishes control-plane and data-plane isolation \cite{27.kubernetes2026multitenancy}. Yet none of these viewpoints alone answer four cross-system questions: which elements are constituents, what remains locally controlled, which contracts are shared or scoped, and what evidence proves that a workflow is intentionally composed rather than accidentally wired together. That gap motivates a platform-level SoS model rather than another universal scheduler or reference topology.

\section{Requirements and Design Principles}

The proposed SoS lens begins with system boundaries, responsibilities, and workflow intent rather than a specific implementation, technology stack, topology, or prescriptive deployment blueprint. The following statements consolidate the requirements and design principles that guide composable integration across independently controlled systems. They draw on SoS engineering, Cloud-HPC convergence, edge-continuum research, security guidance, artifact management, and interoperability work \cite{6.maier1998architecting,7.iso21839,8.iso21840,19.belcastro2026navigating,20.rosendo2022distributed,25.guo2024hpcsecurity,26.rose2020zerotrust,27.kubernetes2026multitenancy,28.wang2009lcim,29.schlegel2022management}. 

\textit{D1: Preserve constituent-system autonomy.} Native schedulers, orchestrators, data systems, managed fabrics, identity services, observability platforms, and edge controllers retain the control that makes them independently useful.

\textit{D2: Make boundaries and ownership explicit.} The SoS view distinguishes components, subsystems, constituent systems, and shared integration services. It also identifies ownership of scheduling, metadata, deployment, identity, telemetry, and recovery. 

\textit{D3: Compose through interfaces and contracts.} Cross-system behavior is coordinated through APIs, events, references, policy mappings, deployment specifications, telemetry, and operational evidence rather than by replacing native control planes. 

\textit{D4: Support multiple execution mappings.} Batch and distributed training, serving, batch inference, stream ingestion, interactive sessions, and edge inference or control may be assigned to, or span, different systems. Training and inference are lifecycle stages rather than fixed constituent-system classes.

\textit{D5: Treat data, artifacts, provenance, and policy as integration objects. }Dataset references, checkpoints, model artifacts, registry entries, lineage, retention, access context, and transfer paths are explicit rather than hidden inside scripts or job directories \cite{29.schlegel2022management}.

\textit{D6: Make tenancy, observability, and recovery cross-system properties.} Isolation should reflect trust relationships and data sensitivity. Telemetry and audit evidence should support diagnosis, accountability, fault containment, and recovery \cite{25.guo2024hpcsecurity,26.rose2020zerotrust,27.kubernetes2026multitenancy,30.opentelemetry2026,31.prometheusOverview}. 

\textit{D7: Remain topology-neutral and evolvable.} Constituent-system boundaries and cross-system contracts should remain independent of a particular physical deployment, virtualization model, network topology, orchestration stack, or vendor implementation. Constituent systems should be added, replaced, or reconfigured through defined interfaces and contracts without redesigning the overall integration model.

\textit{D8: Target selective workflow interoperability.} The objective is sufficient technical, syntactic, semantic, and pragmatic interoperability to carry project, identity, dataset, job, artifact, endpoint, telemetry, quota, and failure context for a selected workflow, rather than universal interoperability among all systems \cite{28.wang2009lcim}.

Together, D1–D8 define complementary views for reasoning about peer constituent-system boundaries, constituent qualification, local, shared, and scoped responsibilities, integration surfaces, workflow realization, and system evolution. These views separate the principal integration concerns while allowing implementations, technologies, and topology to change without altering the underlying SoS reasoning model \cite{33.iso42010}.

\section{System-of-Systems Integration Model}
\subsection{Peer Constituent-System View}
\begin{figure}
    \centering    \includegraphics[width=1\linewidth]{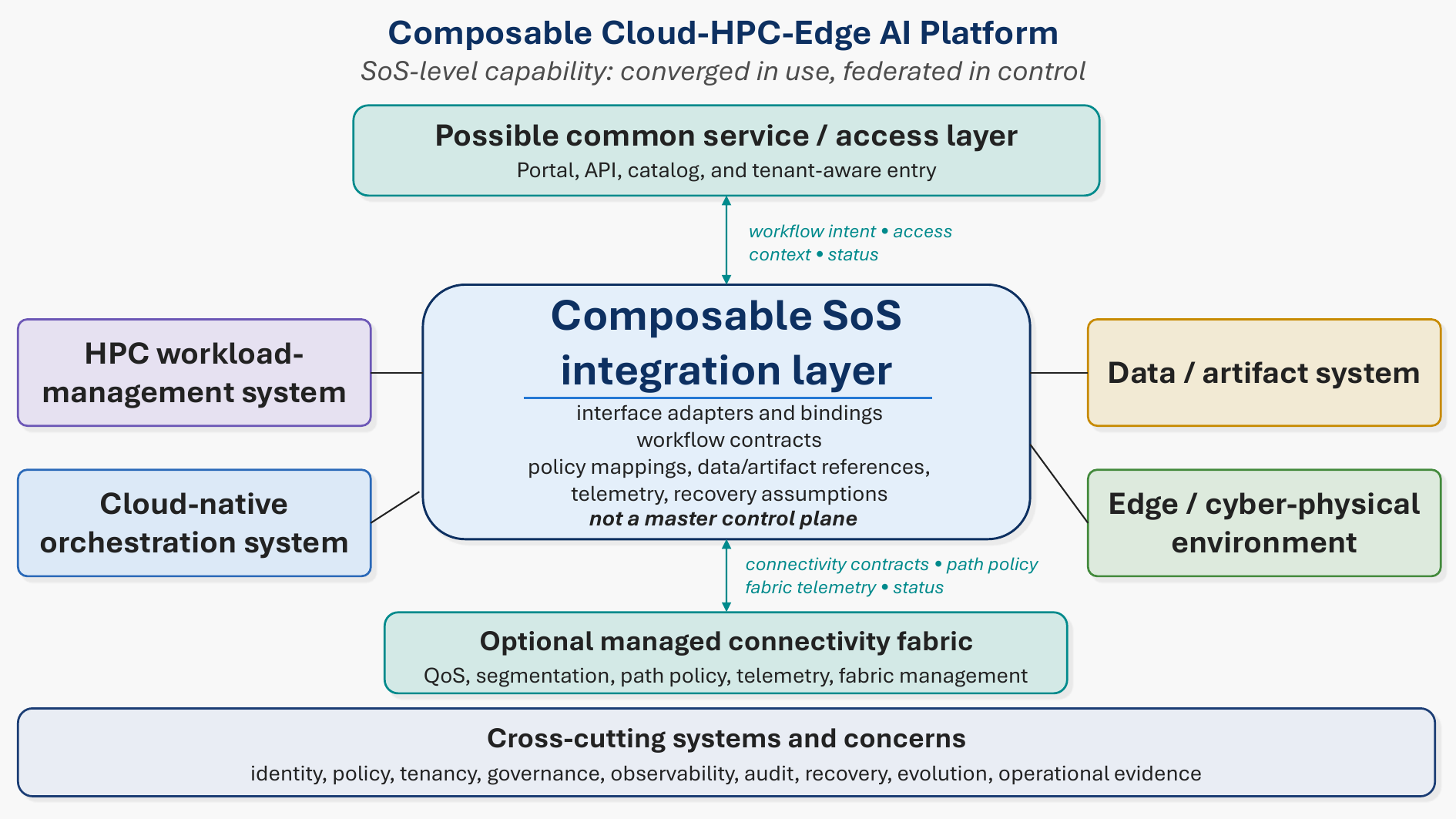}  
     \Description{Figure 1}
    \caption{Peer constituent-system classes integrated through a composable SoS integration layer, with a possible common service/access layer, an optional managed connectivity fabric, and cross-cutting systems and concerns. The constituent-system blocks represent distinct operating and management boundaries rather than fixed workload destinations. The service/access layer and managed connectivity fabric represent optional roles and are not assumed to be constituent systems by default. Constituent systems may interoperate through shared interfaces and contracts while retaining local control.}
    \label{fig:Figure 1}
\end{figure}

Figure \ref{fig:Figure 1} uses four peer constituent-system classes in the base view: an HPC workload-management system, a cloud-native orchestration system, a data/artifact system, and an edge/CPS environment. The classes are peers because each may own a distinct control boundary, operator group, lifecycle, policy model, and failure domain. They represent persistent operating and management boundaries rather than steps in an execution sequence. Training, serving, data preparation, and evaluation may be assigned to one constituent-system class or span several classes.

The peer view, therefore, avoids treating infrastructure as a homogeneous resource pool. The HPC system retains authority over queueing, accounting, allocation, and locality. The cloud-native system retains authority over API state, controllers, namespaces, and service lifecycle. The data/artifact system controls metadata, storage layout, versioning, retention, and lineage. The edge/CPS environment controls devices, gateways, local capture, safety or privacy enforcement, and disconnected operation. The higher-level AI platform capability emerges from their coordinated contribution rather than from transferring these responsibilities to a new controller.

The composable SoS integration layer is therefore a binding role rather than another execution system. It defines adapters, contracts, mappings, references, and evidence needed at system boundaries. It may be implemented by several services or mechanisms and need not lie on every data path. A possible common service/access layer can expose portals, APIs, catalogs, and tenant-aware entry without becoming a constituent by default. A managed connectivity fabric is included only when connectivity has independent management, policy, telemetry, lifecycle, and failure behavior. Cross-cutting concerns such as identity, governance, observability, audit, recovery, and evolution may remain local, be provided through shared services, or be handled by independent systems that satisfy the boundary test themselves.

This separation underlies the principle: \textit{converged in use, federated in control}.  Users may experience a common platform entry point and workflow-status view, while each participating system remains responsible for its local decisions. Shared interfaces, contracts, and operational evidence provide enough context to coordinate workflows and diagnose failures without requiring complete state replication or global control.

Figure \ref{fig:Figure 1} presents a structural view of operating and management boundaries and the integration relationships among them. The four peer constituent-system classes retain their native interfaces and control planes, while the composable SoS integration layer coordinates the cross-system exchange of workflow intent, policy context, dataset and artifact references, deployment specifications, telemetry, and recovery expectations. A possible service/access layer may provide a shared entry point but is not a constituent system by default. Managed connectivity and other supporting or cross-cutting capabilities are classified as constituent systems only when they satisfy the boundary test in Section 4.2.

\subsection{Constituent-System Boundary Test}
Constituent-system classification begins by determining which platform elements have both local independence and an explicit role in the higher-level AI platform capability. Workload managers, orchestration environments, data and artifact systems, connectivity fabrics, identity services, observability platforms, and edge/CPS environments are not constituent systems merely because they are technically complex. A candidate exhibits local independence when it provides a meaningful capability of its own, retains native control, has an independently managed lifecycle, and has a recognizable failure and recovery model. It has an integration role when it exposes interfaces or contracts through which it contributes to the higher-level AI platform capability.

Classification is specific to the SoS scope being examined rather than to a technology category. An HPC workload-management system may qualify when it controls queues, resource allocation, accounting, operational policy, and recovery. A cloud-native orchestration system may qualify when it controls its API, desired-state mechanisms, namespaces, service lifecycle, and upgrades. An edge/CPS environment may qualify when it has a local operational purpose, device or gateway control, privacy or safety constraints, and recognizable disconnected or degraded behavior.

\begin{figure}
    \centering
    \includegraphics[width=1\linewidth]{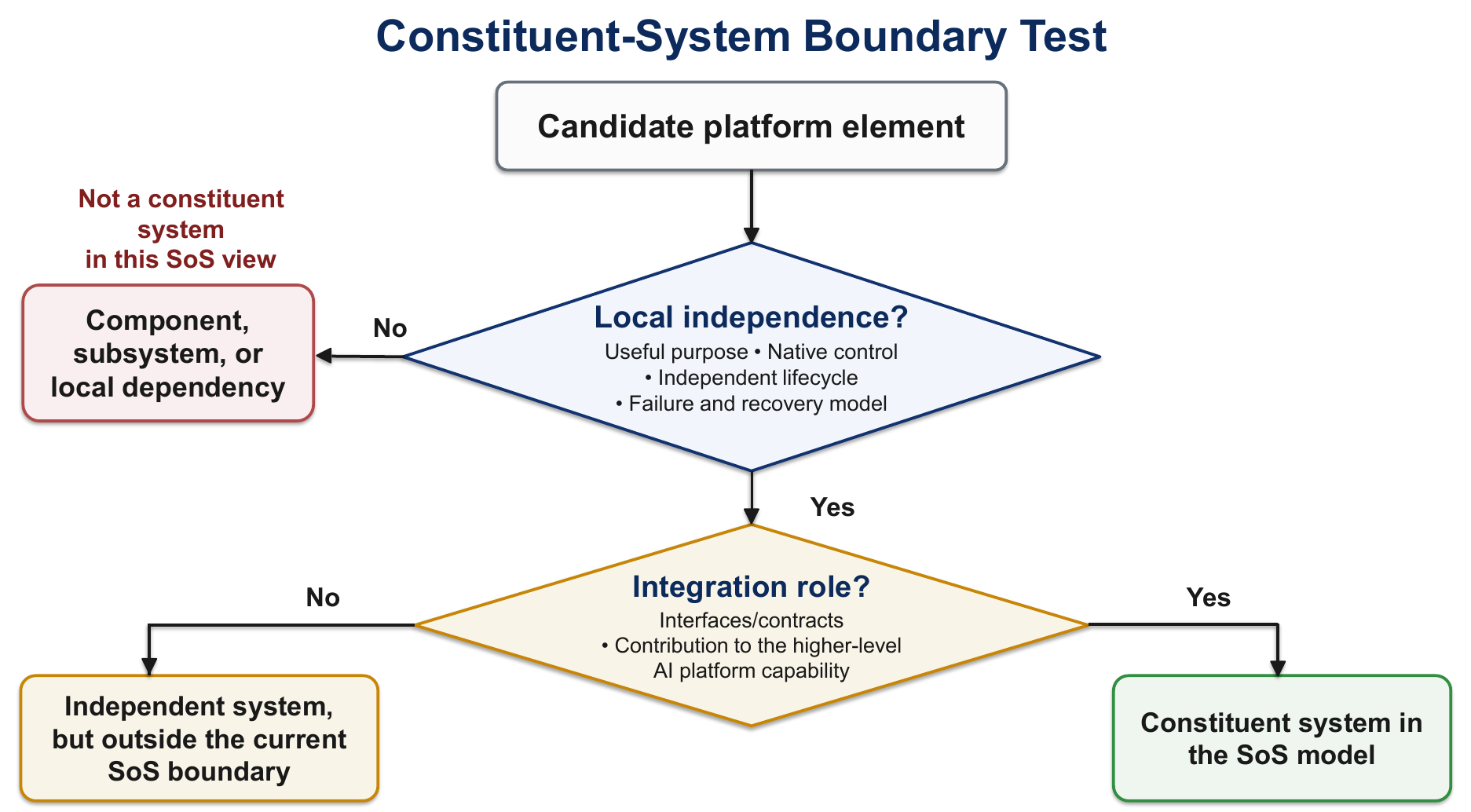}
    \caption{Constituent-system boundary test. A candidate platform element is treated as a constituent system in the current SoS view when it has both local independence and an explicit integration role in the higher-level AI platform capability. An element without local independence is treated as a component, subsystem, or local dependency; an independently useful system without an integration role remains outside the current SoS boundary.}
    \label{fig:Figure 2}
    \Description{Figure 2}
\end{figure}

Capabilities such as data/artifact management, connectivity, identity, and observability are deployment-dependent because they may be operated as part of a larger system or as independent systems. The boundary test shown in Figure \ref{fig:Figure 2} and operationalized by the questions in Table \ref{tab:boundary-test} distinguishes these cases. A capability remains a component, subsystem, or local dependency when it lacks local independence: it does not provide a useful capability of its own or does not retain its own control, lifecycle, and recognizable failure and recovery model. Its scale or technical complexity does not change that classification. For example, a storage pool, network fabric, identity mechanism, or monitoring service operated as part of a larger system would not qualify as a constituent system solely because it is large or technically complex.

Local independence alone is insufficient. An independently useful system may remain outside the current SoS boundary when it does not participate in the higher-level capability. Participation is established through explicit interfaces or contracts, such as job specifications, deployment requests, dataset and artifact references, policy mappings, telemetry context, or recovery expectations. The boundary test, therefore, avoids treating every complex subsystem as a constituent system while recognizing independently controlled systems whose participation materially shapes the operation, governance, recovery, or evolution of the integrated AI platform.

Figure \ref{fig:Figure 2} separates three outcomes. A candidate lacking local independence is treated as a component, subsystem, or local dependency rather than a constituent system in the current SoS view. An independently useful candidate without an explicit integration role in the higher-level AI platform capability remains outside the current SoS boundary. Only a candidate satisfying both conditions is classified as a constituent system. Technical complexity alone therefore does not determine SoS status.

Classification remains context-dependent. A local storage pool, internally managed network, embedded monitoring capability, or system-specific identity mapping is normally treated as a component or subsystem when it is operated as part of a larger system. By contrast, a storage platform, managed WAN or RDMA fabric, identity provider, or observability service may qualify as a constituent system when it has an independent operational purpose, operators, management plane, lifecycle, interfaces, and recovery procedures.

\begin{table}[t]
\centering
\caption{Operational questions for the constituent-system boundary test.}
\label{tab:boundary-test}
\footnotesize
\setlength{\tabcolsep}{3pt}
\renewcommand{\arraystretch}{1.05}

\begin{tabular}{
    p{0.27\columnwidth}
    p{0.65\columnwidth}
}
\hline
\textbf{Boundary question} &
\textbf{Operational interpretation} \\
\hline

Useful purpose &
If platform integration is removed, does the element still provide a meaningful capability? If not, it is probably a component or subsystem. \\

Native control &
Does it have its own scheduler, controller, fabric manager, identity provider, metadata service, or management plane? \\

Independent lifecycle &
Can it be owned, operated, upgraded, staffed, or budgeted separately from the rest of the platform? \\

Failure and recovery model &
Does its failure create a recognizable operational event with an independently defined recovery procedure? \\

Interfaces/contracts &
Does it expose APIs, paths, events, policies, telemetry, or service endpoints through which other systems interact with it? \\

Higher-level contribution &
Does it contribute to the higher-level AI platform capability that no single participating system can provide alone? \\

\hline
\end{tabular}
\end{table}

\subsection{Local, Shared, and Scoped Responsibilities}
The boundary test identifies which systems participate in the SoS; the responsibility model clarifies how authority and coordination are divided among them. Local authority remains with native constituent systems. Shared responsibilities are limited to the contracts and context required for cross-system coordination. Scoped controls apply additional isolation, reservation, or policy to a project, tenant, organization, site, workflow, or risk class. Table \ref{tab:responsibilities} applies this distinction across the principal platform areas.

\begin{table}[t]
\centering
\caption{Local, shared, and scoped responsibilities.}
\label{tab:responsibilities}
\scriptsize
\setlength{\tabcolsep}{2pt}
\renewcommand{\arraystretch}{1.08}

\begin{tabular}{
    p{0.16\columnwidth}
    p{0.26\columnwidth}
    p{0.27\columnwidth}
    p{0.25\columnwidth}
}
\hline
\textbf{Area} &
\textbf{Local authority of constituent systems} &
\textbf{Shared through composable integration} &
\textbf{Scoped or isolated when required} \\
\hline

\textbf{Control and execution} &
Native scheduling, desired-state control, placement, execution, accounting, and local recovery. &
Submission and deployment interfaces, workflow identifiers and state, handoff events, placement constraints, and quota or allocation summaries. &
Project queues or partitions, namespaces, reservations, virtual clusters or control planes, and dedicated edge-control domains. \\

\textbf{Data and artifacts} &
Storage layout, metadata services, file and object stores, checkpoints, model registries, edge buffers, and retention enforcement. &
Dataset and artifact identifiers, locations, versions, provenance, access context, and transfer rules. &
Tenant buckets or volumes, encryption domains, quotas, retention policies, and regulated-data boundaries. \\

\textbf{Connectivity} &
Routing and fabric control, segmentation enforcement, WAN or VPN operation, failure domains, and local network telemetry. &
Reachability requirements, permitted paths, latency or QoS intent, exposure rules, connectivity status, and fabric telemetry. &
Tenant VLANs or VRFs, RDMA domains, dedicated tunnels or WAN paths, and isolated network segments. \\

\textbf{Identity and policy} &
Native identities, service accounts, authorization models, and local policy enforcement. &
Identity federation, subject and role mapping, policy translation, trust context, and audit-correlation identifiers. &
Tenant identity providers, project-specific roles and service accounts, external trust domains, and workflow-specific data-use policies. \\

\textbf{Observability and recovery} &
Local metrics, logs, traces, health checks, alerts, runbooks, and recovery mechanisms. &
Cross-system correlation, workflow telemetry, SLO context, audit trails, and failure or recovery status. &
Restricted dashboards and log views, tenant or project SLOs, regulated audit retention, and workflow-specific recovery objectives. \\

\hline
\end{tabular}
\end{table}

\section{Composable Integration Surfaces}
Composable integration operates at the platform/system-integration layer. It does not pool hardware under a single infrastructure controller, nor does it turn independently managed systems into microservices. Its purpose is to expose enough stable structure for a workflow to cross boundaries while native systems retain authority.

We identify seven primary integration surfaces. The \textit{access surface} defines how users, tenants, and external services submit workflow intent, discover capabilities, and obtain status. It may be realized through a common portal, APIs, command-line or notebook gateways, catalogs, native system entry points, or a combination of these. The \textit{control-plane interface surface} exposes native capabilities through scheduler interfaces, orchestration APIs, storage APIs, fabric telemetry, edge-deployment interfaces, and policy endpoints. The \textit{execution surface} maps requested activities to batch jobs, containers, services, notebooks, pipelines, or edge inference and control tasks. The \textit{data and artifact surface} carries dataset and artifact references, object or file paths, checkpoints, registry entries, model versions, provenance, and transfer rules \cite{29.schlegel2022management}.

The \textit{observability surface} defines how metrics, logs, traces, alerts, SLO context, audit events, and incident identifiers are exposed and correlated across system boundaries. OpenTelemetry and Prometheus provide useful mechanisms \cite{30.opentelemetry2026,31.prometheusOverview}, but the architectural requirement is to preserve workflow, tenant, and provenance context as operational evidence crosses systems. The \textit{policy and governance surface} defines how identity, roles, quotas, tenant isolation, data-access rules, placement constraints, and compliance obligations are translated across local control domains, while enforcement remains with the native systems. The \textit{recovery and evolution surface} makes explicit the failure domains, recovery responsibilities, fallback paths, compatibility constraints, upgrade boundaries, and conditions under which a constituent system may be replaced or extended.

Each integration surface should expose only the information and capabilities required by the workflow. Native constituent systems retain authority over their internal operations: the HPC workload manager controls job scheduling and accounting; the cloud-native orchestrator controls services and desired state; storage systems control metadata and data layout; managed connectivity fabrics, when present, control network behavior; and edge/CPS systems control local execution. The composable integration layer exchanges workflow intent, references, policy context, and operational evidence without taking over these local decisions. This separation of cross-system coordination from native control distinguishes composable SoS integration from a monolithic platform control plane.

\section{Representative Cross-System Workflow}
Figure \ref{fig:Figure 3} illustrates how the integration surfaces operate in a remotely initiated edge-to-training-to-serving feedback loop. The example is illustrative rather than prescriptive. It does not define a mandatory pipeline or permanently assign lifecycle stages to particular constituent-system classes. Instead, it shows how activities remain under native control while cross-system handoffs carry explicit workflow intent, references, policy context, and operational evidence. In other deployments, edge capture may operate autonomously and enter the workflow at dataset registration or another coordination point.

\subsection{Remotely Initiated Edge - to - Training - to - Serving Feedback Loop}

The proposed integration model can support multiple cross-system AI workflows. We use a remotely initiated edge-to-training-to-serving feedback loop because it spans the principal constituent-system classes shown in Figure \ref{fig:Figure 1} and exercises several distinct integration needs. These include workflow intent, edge-side data selection, dataset registration, HPC-managed training or evaluation, artifact publication, cloud-native serving, edge deployment, telemetry, policy context, and feedback for retraining. Other workflows may omit, reorder, replace, or add activities, participating systems, and handoffs.

A user or tenant initiates the workflow through a possible common service/access layer. The submitted workflow or control intent may configure edge-side capture and local filtering, while the edge/CPS environment retains authority over device operation, privacy enforcement, and local data selection. Selected data is then registered in the data/artifact system together with the metadata, policy context, and access information required by downstream activities. The data/artifact system exposes the registered dataset through an explicit reference, which is combined with a job specification and submitted to the HPC workload-management system. The HPC system applies its native queueing, placement, accounting, and failure-handling policies. 

Training, fine-tuning, or evaluation produces a model artifact, an evaluation result, and associated lineage information. These outputs are published through the data/artifact system, which provides the versioned references required by downstream consumers. The published artifact may be used by the cloud-native orchestration system to create a serving endpoint, by the edge/CPS environment for edge deployment, or by both. These are independent downstream mappings rather than sequential stages.

\begin{figure}
    \centering
    \includegraphics[width=1\linewidth]{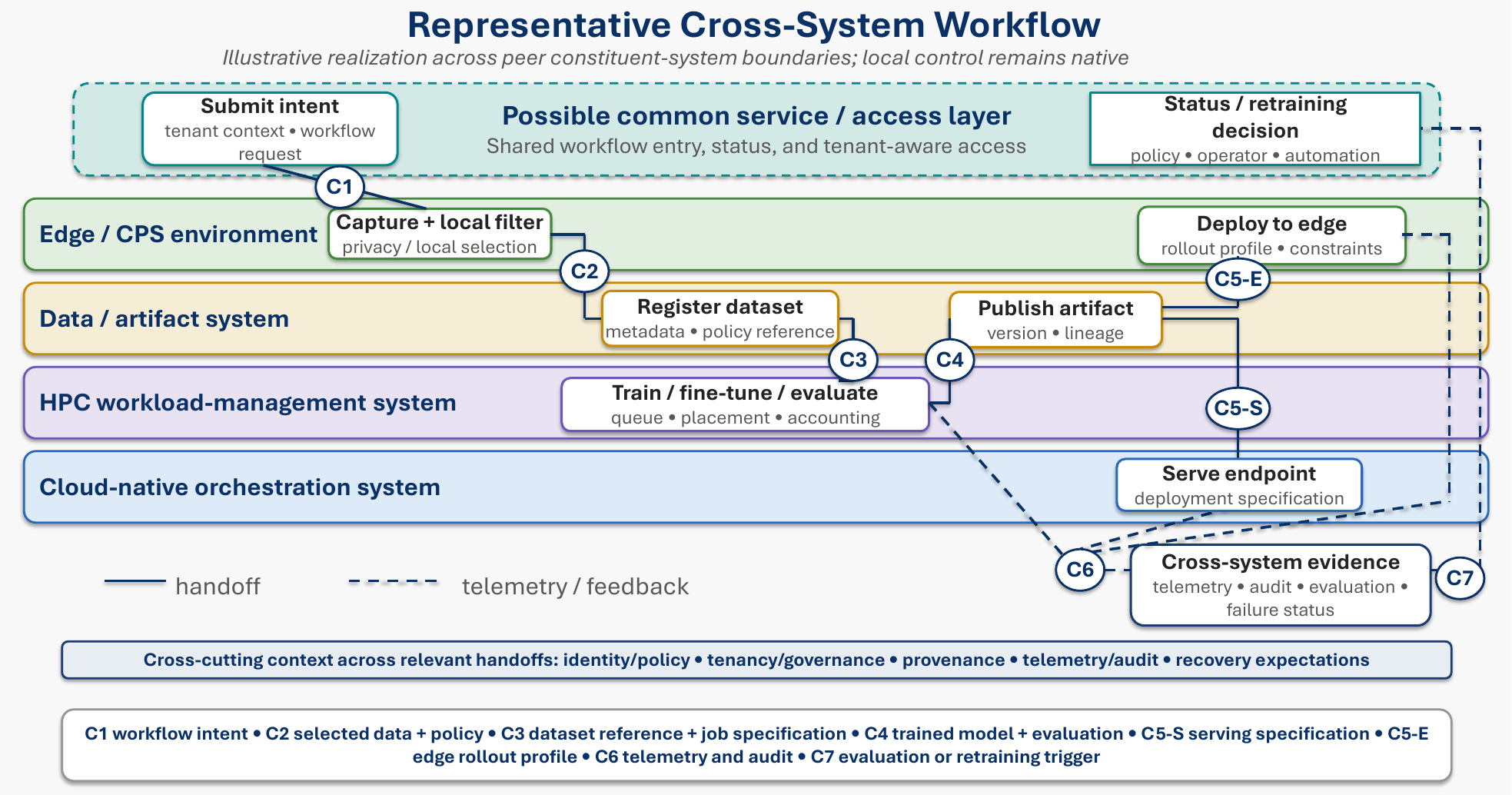}
     \Description{Figure 3}
    \caption{Representative cross-system AI workflow across peer constituent-system boundaries. A possible common service/access layer initiates the remotely triggered workflow, while activities execute under the native control of the edge/CPS, data/artifact, HPC workload-management, and cloud-native orchestration systems. Solid arrows denote governed handoffs; dashed arrows denote telemetry, operational evidence, and feedback. After artifact publication, cloud-native serving and edge deployment are independent downstream destinations. Cross-cutting context accompanies relevant handoffs rather than forming a separate workflow stage.}
    \label{fig:Figure 3}
\end{figure}

The solid arrows in Figure \ref{fig:Figure 3} denote governed workflow handoffs. C1 carries workflow or control intent. C2 carries selected data and policy context. C3 carries the dataset reference and job specification. C4 carries the model artifact and evaluation result. C5-S and C5-E carry the serving specification and edge rollout profile, respectively. The dashed C6 arrows carry telemetry, audit records, evaluation results, and failure evidence into the cross-system evidence view. The dashed C7 path carries status, evaluation outcomes, or an optional retraining trigger back to the service/access layer. Identity and policy, tenancy and governance, provenance, telemetry and audit, and recovery context are attached only to the handoffs for which they are relevant.

\subsection{Alternative Workflow and Tenancy Mappings}
The workflow mapping in Figure \ref{fig:Figure 3} is illustrative rather than fixed. For example, a cloud-native service endpoint may use accelerators allocated by an HPC workload-management system. Prior work integrating Kubernetes, Slurm, and vLLM demonstrates the feasibility of this execution pattern \cite{18.trappen2025automated}. From the SoS perspective, the relevant requirement is that the cross-system interface preserves HPC allocation, accounting, observability, and failure semantics while the service-facing system retains responsibility for endpoint lifecycle.

The same logical workflow may also be scoped differently for different tenants or trust relationships. A multi-tenant research platform may expose common training, storage, serving, and edge capabilities through distinct namespaces, identity mappings, storage domains, network segments, audit views, reservations, or virtual control planes. Kubernetes multi-tenancy guidance and virtual-cluster research show that control-plane and data-plane isolation are architectural choices rather than minor deployment details \cite{27.kubernetes2026multitenancy,32.zheng2021multitenant}. These variants reinforce the central principle: workflow stages may be assigned to or span different constituent-system classes, while native systems retain local authority, and cross-system behavior is coordinated through explicit interfaces and contracts.

\section{Feasibility Evidence and Research Agenda}
A credible SoS integration approach must be testable at the level of system boundaries, workflow handoffs, operations, tenancy, evolution, and cost. We therefore identify six classes of evidence. \textit{Boundary and ownership evidence} tests whether operators can consistently distinguish components, constituent systems, and external systems, and whether ownership, interfaces, and failure domains are explicit. \textit{Workflow and interoperability evidence} tests whether each cross-system handoff can be described at the technical, syntactic, semantic, and pragmatic levels \cite{28.wang2009lcim}. \textit{Operational evidence} tests whether telemetry and audit records preserve workflow context, support diagnosis, and demonstrate fault containment. \textit{Tenancy and policy evidence} tests whether the same workflow contracts can be safely scoped for internal, external, or regulated users. \textit{Evolution and replacement evidence} tests whether a scheduler, data system, edge site, serving runtime, identity provider, or observability backend can be added or replaced without redesigning the full platform. \textit{Performance and operational-cost evidence} tests whether the benefits of composition justify its overhead in latency, throughput, integration effort, diagnosis time, policy administration, and workflow reuse.
These evidence classes lead to five research questions. \textbf{RQ1}: How should operators distinguish a constituent system from a component, subsystem, or external system? \textbf{RQ2}: What is the minimum interface and contract description required for independently controlled systems to participate in reusable workflows? \textbf{RQ3}: How should shared concepts such as project, job, dataset, model artifact, tenant, quota, SLO, provenance, and failure domain be mapped across control planes? \textbf{RQ4}: How should placement and handoff decisions account for locality, latency, accelerator type, energy, isolation, and recovery, building on continuum-scheduling work such as DECICE \cite{21.sharma2026decice}? \textbf{RQ5}: How can operators diagnose cross-system failures, verify containment, and replace a constituent system while preserving workflow contracts and operational evidence?
Evaluation can proceed incrementally. An initial prototype should classify the participating systems, execute one governed cross-system workflow, and correlate telemetry across its handoffs. A stronger evaluation should add fault injection, tenant-specific policies, constituent-system replacement, and measurements of interface overhead, diagnosis effort, and workflow reuse.

\section{Conclusion}
This paper advances a system-of-systems framing for composable Cloud-HPC-Edge AI platforms. It makes constituent-system boundaries, ownership, interfaces, contracts, and operational evidence explicit. Consistent with the architecture-description perspective of ISO/IEC/IEEE 42010 \cite{33.iso42010}, the framing is expressed through complementary views of peer constituent-system classes, constituent qualification, local, shared, and scoped responsibilities, integration surfaces, workflow realization, and system evolution.

Cloud-HPC-Edge AI platforms become difficult to reuse when portals, schedulers, orchestrators, data services, connectivity fabrics, observability mechanisms, and edge environments are connected through deployment-specific scripts, implicit mappings, and undocumented failure assumptions. A system-of-systems perspective provides a more disciplined alternative. Composable integration preserves native authority while defining reusable contracts for access, control-plane interaction, execution, data and artifacts, policy and governance, observability, recovery, and evolution.

The proposed direction is converged in use, federated in control, and composed through explicit interfaces and evidence. This framing can serve as a basis for reproducible architecture descriptions, testable evaluation criteria, incremental constituent-system replacement, and future federation across sites and organizations. It may also extend to additional independently managed systems as computing platforms evolve. For example, emerging hybrid quantum-HPC environments introduce distinct execution models, control mechanisms, interfaces, and lifecycles that could be incorporated using the same boundary and integration principles \cite{36.eurohpc2026quantum}.

\section*{Disclosure}
Generative AI tools assisted with language editing, graphics enrichment, and reference checking. The authors verified the claims, citations, and final wording and take responsibility for the content.

\bibliographystyle{ACM-Reference-Format}
\bibliography{socc_vision_refs}
\end{document}